**Incommensurate crystal supercell and polarization-flop observed in the magnetoelectric ilmenite, MnTiO$_3$**

Harlyn J. Silverstein[1,2], Elizabeth Skoropata[3], Paul M. Sarte[1], Cole Mauws[4], Adam A. Aczel[5], Eun Sang Choi[6], Johan van Lierop[3], Christopher R. Wiebe[1,4,7,8] and Haidong Zhou[6,9]

1 – Department of Chemistry, University of Manitoba, Winnipeg R3T 2N2, Canada
2 – Department of Applied Physics, Stanford University, Stanford, California 94305-4045, United States
3 – Department of Physics and Astronomy, University of Manitoba, Winnipeg R3T 2N2, Canada
4 – Department of Chemistry, University of Winnipeg, Winnipeg R3B 2E9, Canada
5 – Quantum Condensed Matter Division, Oak Ridge National Laboratory, Oak Ridge 37831-6475, United State
6 – National High Magnetic Field Laboratory, Florida State University, Tallahassee 32306-4005, United States
7 – Department of Physics and Astronomy, McMaster University, Hamilton L8S 4M1, Canada
8 – Canadian Institute for Advanced Research, Toronto M5G 1Z7, Canada
9 – Department of Physics and Astronomy, University of Tennessee-Knoxville, Knoxville 37996-1220, United States

MnTiO$_3$ has been studied for many decades, but it was only in the last few years that its strong magnetoelectric behavior had been observed. Here, we use neutron scattering on two separately grown single crystals and two powder samples to show the presence of a supercell that breaks R-3 symmetry. We also present the temperature and field dependence of the dielectric constant and pyroelectric current, and show evidence of non-zero off-diagonal magnetoelectric tensor elements (forbidden by R-3 symmetry) followed by a polarization flop accompanying the spin flop transition at $\mu_0 H_{SF}$ = 6.5 T. Mössbauer spectroscopy on MnTiO$_3$ gently doped with $^{57}$Fe was used to help shed light on the impact of the supercell on the observed behavior. Although the full supercell structure could not be solved at this time due to a lack of visible reflections, the full-scope of the results presented here suggest that the role of local spin-lattice coupling in the magnetoelectric properties of MnTiO$_3$ is likely more important than previously thought.

**1.0 – Introduction**

Over the past century, the field of condensed matter has evolved from a small group of scientists looking to understand simple materials into the single largest branch of physics being studied today. Massive collaborations have resulted in the construction of various national and international facilities with the sole purpose of finding, preparing, and tailoring new materials. This has largely occurred out of necessity as new-found systems become evermore complex. Still, old materials that were thought to be well understood can yet hold secrets that are waiting to be discovered.

Such is the case with MnTiO$_3$; an ilmenite material whose crystal structure motif was first described nearly 80 years ago. [1, 2] Containing a layered, rhombohedral structure, Mn$^{2+}$ moments couple antiferromagnetically along the hexagonal setting *c*-axis (for the duration of this work, the hexagonal setting will be used). Between the 1950s and 1990s, the magnetic properties of MnTiO$_3$ were studied using magnetometry [3, 4], neutron scattering [5-8], heat capacity [9], and electron paramagnetic resonance. [10] A field applied along this easy axis results in a spin-flop transition into the *ab*-plane near $\mu_0H_{SF}$ = 6.5 T. [11] $MnTiO_3$ exhibits quasi-two dimensional behaviour resulting from an accidental cancellation of various, otherwise strong, interplanar exchange interactions. [6, 7, 12, 13] This is in contrast to other 3*d* ilmenites such as $FeTiO_3$, $CoTiO_3$, and $NiTiO_3$, but the difference in behaviour is largely accounted for by a modification of the Goodenough-Kanamori rules [14] using an exciton-superexchange mechanism. [15]

But, over the last eight years $MnTiO_3$ has once again been the subject of scientific scrutiny for its strong magnetoelectric properties. [16-22] Mufti *et al.* [16] showed that the application of electric and magnetic fields along the *c*-axis results in a strong dielectric anomaly and pyroelectric current up until the spin flop transition. They used a simple Landau free energy argument to show how the magnetoelectric effect arises. However, doping with even a small amount of $Ni^{2+}$ into the $Mn^{2+}$ site substantially changes both the magnetic and electrical properties of the system [18, 20, 22], which complicates the mechanism for magnetoelectric behaviour.

In this report, we present a study of the magnetic and electric properties of both powder and single crystal $MnTiO_3$. Using neutron diffraction, we report the existence of nuclear supercell reflections found in four independent samples (both powder and single crystal) using two separate instruments. In addition, we elaborate on our previous dielectric and pyroelectric measurements [20] that show the existence of an electric polarization flop beyond the spin flop transition temperature, which is very rare for materials containing only 3*d* magnetic elements. Mössbauer spectroscopy on a 1 wt% $^{57}$Fe-doped sample shows evidence of the influence of the supercell at the local level. Finally, we discuss the potential role of the supercell regarding the properties of this material in the context of strong spin-lattice coupling and discuss a course for future experimentation.

**2.0 – Experimental**

Stoichiometric amounts of MnO and $TiO_2$ were combined and ground by hand for 15 minutes per gram of reactants using an agate mortar and pestle. The ground material was pelleted using a hydrostatic rod press and placed in a tube furnace using an uncovered alumina crucible. The temperature was ramped to 800°C at five degrees per minute and left for 30 minutes. This was followed by sintering at 1350°C for five hours. An annealing step was implemented at 1150°C for two days in enriched $O_{2(g)}$ with intermediate re-grinding and pelleting. Another sample was prepared in air doping 1 wt% $^{57}$Fe (97% enrichment), which we call the Fe-sample. Two single crystals were grown using an IR Image Furnace (Canon) with 10 g of either the stoichiometric sample or the Fe-sample in an $O_{2(g)}$ environment. Both single crystals were annealed in a furnace in $O_{2(g)}$ prior to measurements for 72 hours.

To characterize our samples, electron microprobe, DC-susceptibility, AC magnetometry, heat capacity, and low temperature x-ray diffraction were performed (the results of which we present in the supplementary information since the general features χ, $T_N$=64 K, and temperature dependence of the lattice constants are all consistent with previous studies). [3-10, 23] X-ray diffraction measurements were performed on a Brüker D8 Advanced DaVinci diffractometer equipped with Cu Kα source, Ni filter, and Lynxeye detector. Powder neutron diffraction was carried out on the HB-2A diffractometer at the High Flux Isotope Reactor (HFIR, Oak Ridge, TN). [24] For each experiment on the stoichiometric and Fe-sample, five grams of sample were loaded into vanadium cans. Diffraction patterns were collected at both 300 K and 10 K using a wavelength of 1.54 Å. Joint x-ray/neutron refinements were performed at 300 K. All refinements were done using the FullProf Suite. [25] The background, peak profiles, lattice parameters, atomic coordinates, isotropic thermal parameters, and site occupancies were refined (with special attention paid to parameters that may be correlated). The allowed magnetic irreducible representations used in the low temperature neutron diffraction refinements were found using SARA$h$ Representational Analysis and SARA$h$ refine packages. [26-28]

Single crystal neutron diffraction was carried out using the HB-1A fixed energy triple-axis spectrometer ($E_i$ = 14.6 meV, HFIR, Oak Ridge, TN). Two separate experiments were performed. The stoichiometric crystal was aligned in the HK0 plane using a Laue x-ray camera and checked with the CG-1B instrument at HFIR. The sample was placed on a special aluminum mount inside an eight Tesla vertical magnet (the field aligned along the $c$-axis). Temperatures as low as 1.5 K could be accessed using this setup. Due to the absence of a tilting axis within the magnet, the crystal was misaligned by an estimated 0.8° out of the scattering plane. The same stoichiometric powder sample (in the same vanadium can) used in the HB-2A experiments was also placed in the magnet to check for additional magnetic Bragg reflections at high fields. Here, we stress that the powders were loaded loose in the cans: moving grains and the isotropic nature of the samples prevent us from assigning any meaningful value into the magnetic peak intensities (such as domain selection or a change in irreducible representation). Rather, these quick experiments were used as a crude check to rule out the formation of any *new* crystal or magnetic Bragg reflections or peak splitting arising from the field that may have been inaccessible in the chosen plane of scattering. Such data would have unambiguously signified a field-induced first-order phase transition instead of a spin-flop. These same experiments were repeated for the Fe-sample.

For dielectric and pyroelectric measurements, gold electrodes were sputtered onto the polished thin plate-like samples. Dielectric measurements were made by measuring the capacitance using either a GR1620 AP or AH2700 (Andeen-Hagerling) capacitance bridge at frequencies between 1 and 20 kHz. The samples had surface areas typically between 2 and 6 mm$^2$ and thicknesses of 0.1 to 0.3 mm; for analysis they were approximated as infinite parallel plate capacitors. The pyroelectric current was measured with a Keithly 6517A electrometer when warming up the sample at typically 10 K/min at zero electric field. The sample was cooled under a static poling electric field ($E_{Pol}$ = 15 kV/cm). The polarization was obtained by integrating the pyroelectric current signal as a function of time. All dielectric and polarization measurements were made at the National High Magnetic Field Laboratory (NHMFL, Tallahassee, FL). Transmission Mössbauer spectra were collected with a 10 GBq $^{57}$Co**Rh** source on a WissEl spectrometer in constant acceleration mode. The source drive was calibrated at room temperature with a 6 μm thick α-Fe foil. All spectra were collected with a Janis SHI-850 closed cycle refrigeration system between 10 and 275 K using the powdered crystal Fe-sample.

### 3.0 – Results

X-ray diffraction was performed at temperatures between 300 and 15 K. [23] Within the selected temperature range, the sensitivity of the probe and resolution of the instrument, no symmetry breaking transitions were observed (Figure S1). All reflections could be indexed to the ilmenite structure, and no impurities were found. Powder neutron diffraction was performed at the HB-2A beam line at HFIR between 300 and 10 K on the same stoichiometric MnTiO$_3$ sample. A joint x-ray and neutron refinement of the patterns is presented in Figure 1, while Table 1 lists selected refined parameters, bond valence sums, and agreement numbers. We note that all occupied sites were assumed to be 100% occupied, which resulted in excellent fits to the data. Refining the occupancies did not improve the fits, and in some cases, made the fits worse.

Ilmenites are ordered derivatives of the corundum structure (space group R-3c): slightly distorted close-packing of $O^{2-}$ ions with cations occupying 2/3 of the octahedral holes (Figures 2a and b). The remaining vacancies are situated so as to minimize the cationic electrostatic repulsion. Each cation has four neighbours: one along the *c*-axis and three within the *ab*-plane. If the lone metal cation is replaced with two cations such as $Mn^{2+}$ (crystal radius of 0.97 Å) and $Ti^{4+}$ (0.745 Å) [29] in an ordered, layered arrangement, the system loses its *c*-glide plane, resulting in a system that adopts the lower symmetry R-3 space group. Both $Mn^{2+}$ and $Ti^{4+}$ are octahedrally coordinated with oxygen with a slight rhombic distortion. $Mn^{2+}$ and $Ti^{4+}$ cations are arranged in hexagonal layers along the *ab*-plane. The peculiar stacking arrangement at any particular site corresponds to a $Mn^{2+}$-$Ti^{4+}$-□-$Ti^{4+}$-$Mn^{2+}$-□ order travelling up along [001]. Electrostatic repulsion between neighbouring cations along [001] causes a slight buckling of the hexagonal layers (Figure 2c). Although an electron microprobe can be used to identify deviations in cationic stoichiometry [23], determining oxygen stoichiometry to within less than 5%, even through the use of methods like thermogravimetry, can be more difficult. In order to ensure nominal stoichiometry, large sample batches (>20 grams) were prepared to minimize deviations due to weighing errors, and samples were annealed for long periods of time in an oxygen-rich environment. In any case, one of the best determinants of crystal stoichiometry in magnetic oxides is the value of the Néel temperature, $T_N$, which is highly sensitive to such deviations.

By 10 K, the system has gone through an antiferromagnetic transition (Figure 1b). SARA*h* was used to find the possible irreducible representations of the magnetic structure. [23] The observed magnetic reflections were consistent with a **k** = (0, 0, 0) magnetic propagation vector with the R-3 space group. Irreducible representations associated with the space group and propagation vector are limited to collinear ordering of the $Mn^{2+}$ moments. Γ(2) yielded the best fit to the data, which corresponds to the moments ordering in an antiparallel fashion along the *c*-axis, agreeing well with previous neutron scattering work. [5]

We used HB-1A to investigate the effects of an applied magnetic field along the easy axis as the spin-flop transition occurs. Here, an 8 T vertical magnet was used, restricting our scattering vector to lie in the HK0 plane. A thorough mesh scan was performed by systematically scanning all of reciprocal space kinematically accessible by the instrument. Two small peak-like features were found flanking either side of the (110) reflection (Figure 3a). The intensities of these features are less than 1% of the (110) peak, and occur at about 0.03 reciprocal lattice units (r.l.u.) along HH0 and at approximately ±0.1 r.l.u along -KK0. These features do not correspond to an aluminum powder ring or to any position that may arise from incomplete filtering of λ/2 neutrons. A small mesh scan was conducted surrounding the (2-10) reflection, which appears at the same |**Q**| as (110) and represent the (1-10) and (10-1) reflections in the rhombohedral setting respectively. Again, extra reflections were observed (Figure 3b). Radial scans were performed on these peaks with a representative shown in Figure 3c.

The presence of these additional features is surprising and would seem to suggest an incommensurate ordering of the magnetic moments along the HK0 plane above the spin flop transition.

The same mesh scan was repeated with $\mu_0H=0$ T. Again, the same extra reflections were observed, with neither the peak intensities nor the peak positions having changed as a function of field (Figure 3c). Even when the temperature was raised to 300 K, the features remained. The results thus far imply that these features are either instrumental, unique to our crystal, or are general features of MnTiO$_3$ that were not observed in previous studies. The powder averaged position of the peaks in reciprocal space occurs at Q = 2.53 Å$^{-1}$. A weak feature is even found in the stoichiometric *powder* neutron diffraction data at Q = 2.52 Å$^{-1}$ [23], which was assumed to be part of the background scatter upon our first examination, and thus rules out an instrumental origin, impurity, grain or twinning effects. The offset in the positions of the extra reflections from the centres of the (110) and (2-10) reflections are not uncommon, but result from a multi-dimensional supercell propagation vector, similar to that found in nepheline. [30] The extra reflections are completely unobservable in the laboratory x-ray diffractograms. Thus, three conclusions can be drawn from the measurements: 1) the highly penetrating nature of a neutron diffraction experiment indicates that the supercell is a bulk feature rather than a surface effect; 2) the supercell is likely due to incommensurate oxygen, rather than cationic, displacements; and 3) the supercell is likely a general feature of all highly crystalline MnTiO$_3$ samples prepared using the ceramic method.

The magnetoelectric properties of MnTiO$_3$ were examined at the National High Magnetic Field Laboratory up to fields of 35 T (same sample as in [17, 20]). The dielectric constant, $\varepsilon_{cc}$, was first measured by applying both the electric and magnetic fields along the *c*-axis. As reported earlier by Mufti *et al.* [16], no anomaly exists in the absence of a field, even at T$_N$ (Figure S5). Once a magnetic field is applied, a strong and sharp peak begins to form at T$_N$ that not only increases in intensity with applied field up to $\mu_0H$, but also moves to slightly lower temperatures, which is in agreement with Mufti *et al*. [16, 23] A broad shoulder also develops extending from T$_N$ down to 30 K as the applied field reaches the spin flop transition. As the field is increased beyond $\mu_0H_{SF}$, the peak begins to decrease in intensity and eventually merges with the shoulder feature at 8 T (Figure 4a). Beyond 8 T, the single feature decreases in intensity until only a slight kink remains. Due to the decreasing stability of the magnetic field beyond 25 T, the uncertainty associated with the dielectric constant becomes larger; only qualitative trends such as the general temperature dependence of the dielectric constant is relevant.

The dielectric constant, $\varepsilon_{ac}$, was measured by applying the electric field along the *a*-axis while keeping the magnetic field applied along *c*. Up until $\mu_0H_{SF}$, there was no clear influence of an increasing magnetic field on $\varepsilon_{ac}$, but once $\mu_0H \approx \mu_0H_{SF}$ a step-like anomaly developed at 20 K (Figure 4b). At 7 T, the anomaly occurs at 45 K. By 10 T, the anomaly occurs at T$_N$ and remains with higher fields, growing more prominent as the field is further increased (Figure 4c). The change in $\varepsilon_{ac}$ reaches as high as 23% at 35 T. [23] The dielectric constant, $\varepsilon_{aa}$, was measured when both the electric and magnetic fields were applied along the *a*-axis and no features were observed up to 18 T (the maximum field for this measurement, not shown).

The magnetoelectric effect in MnTiO$_3$ has been modelled without the use of a supercell by Mufti *et al.* [16] To check whether or not the supercell has any direct influence at the bulk level, the other off-diagonal component of the dielectric tensor, $\varepsilon_{ca}$, found by applying an electric field along *c* and a magnetic field along the *a*-axis, was measured (Figure 4d). A sharp feature was also found at T$_N$ at all fields measured greater than 1 T, with little change in intensity as a function of field beyond 3 T. While the anomaly at $T_N$ in fields above $\mu_0H_{SF}$ can be explained by a symmetry loss accompanying the spin flop transition, the anomaly in fields below $\mu_0H_{SF}$ is quite unexpected; R-3' symmetry only permits $\varepsilon_{ac} = -\varepsilon_{ca} = 0$, which is not the case here. There are two possibilities that can cause such behaviour: 1) the crystal is grossly misaligned or 2) the symmetry is not R-3'. It is noted that Mufti *et al.* [16] reported no anomalies along the off-diagonal components, directly contrasting with the behaviour observed here. Since the anomaly is very weakly field dependent above 3 T all the way to 18 T, this would be inconsistent with a portion of the signal arising from some other tensor component due to a misalignment because the dielectric dramatically signal changes beyond the spin flop transition. It is also noted that the spin flop transition is not expected theoretically [31, 32] or observed empirically with the magnetic field applied along *a* and the results in Figure 4d are consistent with this, suggesting that the feature is real. However, while a small misalignment is always inevitable, we have presented compelling evidence for supercell reflections that could indicate the loss of R-3' symmetry in $MnTiO_3$ in the bulk, long-range limit. Furthermore, we have shown that the supercell reflections appear to be temperature and magnetic field-independent within the resolution of HB-1A. We also note that the dielectric anomaly is quite weak, rather sharp, and only becomes noticeable at fields greater or equal to 1 T. Our applied electric field was five times larger than that used in Mufti *et al.* [16], which naturally results in a better signal to noise ratio; it is possible that the anomaly was simply missed in previous studies.

A reversible and spontaneous polarization develops at $T_N$ with an applied magnetic field. The polarization increases with both fields applied along the *c*-axis until $\mu_0H_{SF}$, and then begins to decrease to 0 at 8 T. A polarization flop occurs beyond $\mu_0H_{SF}$ as the field is increased further (this data has been presented before by two of us in [20] and we reproduce it here for convenience in Figure 5). This polarization flop is exceedingly rare in materials that lack rare-earth elements, but has been observed before in $MnWO_4$. [33-36] Recently, two of us (HDZ and ESC) had shown that doping $Ni^{2+}$ into $MnTiO_3$ can alter the magnetic field in which the polarization flop occurs. [20] The magnetoelectric coefficient tensor, $\alpha_{ij}$ can be found from $P_i = \alpha_{ij}H_j$. In this case for $\mu_0H < \mu_0H_{SF}$, $\alpha_{cc} = 5.1\pm0.4 \times 10^{-5}$ CGS units, agreeing well with previous measurements. [16, 20] Above the spin-flop transition field, an off-diagonal component of the magnetoelectric tensor develops and takes on a value of $\alpha_{ac} = 4.44\pm0.03 \times 10^{-5}$ CGS units.

To ascertain the local effects that this supercell might have on the magnetism, transmission Mössbauer spectra were collected at various temperatures on a 1 wt% $^{57}$Fe-doped $MnTiO_3$ sample. Before introducing the results, it is important to mention that even a small amount of dopant ions can dramatically influence the structure and properties of a material. To ensure that the dopant Fe ions truly behave as spectator ions with minimal influence over the intrinsic properties of $MnTiO_3$, we have repeated every measurement mentioned thus far – including *all* neutron scattering measurements on both HB-1A and HB-2A – on both powder and single crystal doped samples appropriately. The Fe-doped samples are consistent with the stoichiometric samples in every way, except for a slight decrease in $T_N$, which was observed at 62 K. We will show that the Fe ions are solely trivalent (S=5/2): this is especially important since $FeTiO_3$ is a very common ilmenite [5] with Fe in its divalent state. Joint Rietveld refinements showed no presence of Fe-O impurities and indicated that both Mn and Ti sites were occupied equally within error. No presence of Fe-O impurities were found in any susceptibility or magnetometry measurements, which indicates that Fe is sufficiently incorporated into the bulk of the sample rather than at the sample surface where oxidation is likely. Finally, and most importantly, the supercell reflections found in the stoichiometric samples are still present in the doped samples. [23]

Transmission Mössbauer spectroscopy was performed on a powdered single crystal of the Fe-containing compound. Spectra were collected at temperatures between 10 and 300 K in order to track the temperature dependence of the hyperfine parameters. The fitted spectra are shown in Figure 6. [23] Immediately apparent is the complexity of the spectra at low temperatures, but the high temperature spectra show some anomalous features as well. In particular, significant relaxation manifested in a large site linewidth is observed at all temperatures, the minimum being $\Gamma=0.27(3)$ mm/s at 10 K (compared to the source's natural linewidth of $\Gamma_{nat}=0.133(2)$ mm/s). At 275 K the spectrum is described by two sites in roughly equal proportion, consistent with the finding from powder diffraction that Fe occupies both the Mn and Ti sites. A quadrupolar splitting, $\Delta=0.64(6)$ mm/s, is present due to the rhombohedral distortion of the electric field gradient in the Fe-O octahedra at each site. The isomer shifts of the major sites are $\delta=0.51(2)$ mm/s at 10 K and are consistent with Fe in its trivalent, rather than a divalent state. [37] Unlike other Mössbauer spectroscopy studies on $MnTiO_3$, the consistent fits to the temperature dependent spectra were possible keeping $\Delta$ and $\delta$ constant over the temperature regimes before and after $T_N$ in agreement with the lack of any observable trends in the metal-oxygen bond distances with temperature in the Rietveld refinements. [23]

As the temperature is lowered towards 75 K from 300 K, magnetic relaxation is manifested by way of significant spectral broadening. This is due to the onset of short-range magnetic order above $T_N$, which also indirectly manifests in the magnetometry data and is consistent with earlier studies. [38] More subspectra (corresponding to extra Fe sites or environments) with different hyperfine fields, $B_{HF}$, were needed to fit the spectra at lower temperatures. By 10 K, the hyperfine fields reach $B_{HF}=50.3(1)$ T and 29.4(2) T, respectively for the two major sites. In addition, $B_{HF}(T)$ for one site is well described by a $S=5/2$ Brillouin function (Figure 7a), which also suggests the trivalent nature of the Fe sites. The Brillouin function was used in an attempt to fit the temperature dependence of the hyperfine field for the second site. The best fit came about from using an $S=1/2$ Brillouin function, but this result is unphysical; more likely, the magnetic behaviour at the second site cannot be modelled using the mean-field approximation.

At least two additional sites (for a total of four sites), were required to describe completely the spectra between 10 and 70 K. These sites cannot be described by any known magnetic relaxation model, (e.g. see Ref. [39] and references therein). The nature of these two extra sites account for about 10% of the total spectral area. The minor 10% spectral component is described by a sextet with a monotonic $B_{HF}(T)$ of 18.2(1) T until 55 K, when it collapses to 0, while the other component was fit with a doublet without a hyperfine field over the entire temperature range. The other fitted parameters are $\delta=0.19(1)$ mm/s and $\Delta=0.20(1)$ mm/s, and a quadrupolar doublet with $\Delta=1.75(7)$ mm/s and $\delta=0.27(3)$ mm/s. Because the ilmenite structure is a nearly close-packed network of $O^{2-}$ with cations occupying 2/3 of the octahedral vacancies, it is possible that at such low doping concentrations of $Fe^{3+}$ some of the ions could be occupying additional vacancies. However, $MnTiO_3$ is already electrostatically strained as evidenced by the buckling of the Mn and Ti hexagonal layers. Fe occupying a vacancy position would likely be quite destabilizing for the lattice.  Another possibility that could describe this minor spectral component is the presence of surface impurities, particularly $Fe_2O_3$ clusters that might have reduced $B_{HF}$ and suffer distorted coordination that would provide a Δ compared to that of bulk $Fe_2O_3$. This can likely be ruled out on the basis that no such impurity phases were detected in the diffraction, magnetometry, or heat capacity experiments.  It is possible that compositional surface impurities are undetectable in low quantities, similar to other compounds such as the half-heusler TiNiSn. [40] However, electron microprobe measurements used to examine the surface of a portion of the Fe-containing crystal presented no detectable compositional defects away from the nominal stoichiometry along the bulk of the crystal, except at regions where the crystal was cut along the edges, where the stoichiometry could not be accurately determined. [23] We note that while the Fe-containing crystal was polished before the electron microprobe measurement, the single crystal portion selected to be used for Mössbauer spectroscopy was also polished before powdering.

To help identify whether or not the magnetic relaxation via spin dynamics or diffusion is responsible for the observed temperature dependence of the Mössbauer spectra, temperature and frequency dependent AC-susceptibility on the powdered Fe-doped sample was collected (Figure S3). [23] For all measured frequencies (up to 1 kHz), the real component of the measured susceptibility mirrors the behaviour observed with DC-susceptibility and the imaginary component is temperature independent and significantly weaker.  This is consistent with the behaviour of magnetic insulators with no Eddy currents and a static ground state, within the time scale of the probe.  Previous polarized inelastic neutron spectroscopy studies showed the presence of magnons that are highly influenced by a magnetic field. [8] The timescale of Mössbauer spectroscopy in relation to the hyperfine-field interaction falls in between AC-magnetometry and inelastic neutron scattering [37], so it is possible that what is being observed is due to an inelastic phenomenon that is diffusion based.

Further evidence of this behaviour is provided by the temperature dependence of the relative *f*-factor, which is found through the relative spectral absorption, Figure 7b. The *f*-factor is related to the integral over the phonon spectrum of the first Brillouin zone [37, 38, 41, 42] and *f*(T) is related directly to the Debye temperature, $θ_D$, which was found to be 350(10) K (fitted line in Figure 7b). This $θ_D$ agrees with previous Mössbauer studies and the weighted-average Debye temperature found using heat capacity measurements. [9, 38]  A gradual increase in the *f*-factor of a Mössbauer nucleus should occur as the temperature is lowered provided there are no dramatic changes in the electronic or crystal structures of the material.  However for $MnTiO_3$, an abrupt increase is observed occurring at ~75 K. Such abrupt changes in the *f*-factor in the vicinity of $T_N$ have been observed in a number of materials possessing a wide-range of functional behaviours including spin-phonon coupling in $GaFeO_3$. [42]

**4.0 – Discussion**

The observation of supercell reflections in $MnTiO_3$ is surprising, but perhaps not entirely unexpected; recent synchrotron x-ray experiments show evidence of small monoclinic distortions in both $Fe_2O_3$ and $Cr_2O_3$ away from R-3*c* symmetry. [43] The monoclinic distortion can be used to explain the odd magnetic order in these materials that lacks a three-fold axis. Similarly, a small distortion away from R-3 in $MnTiO_3$ could be used to explain the existence of the weak anomaly at $T_N$ in $ε_{ca}$ but not $ε_{ac}$, the pre-edge feature observed in the O-K edge found using electron energy loss spectroscopy that was not seen in other ilmenites [44], and could also account for all of the unexpected features found in the second harmonic generation data in the paramagnetic and high-field regimes. [17]

This raises the question as to the nature of the supercell. Unfortunately, it was not possible for us to completely index the cell for numerous reasons, the primary one being that more reflections are required. In the neutron powder diffraction data, it is difficult to differentiate between small variations in the background scatter and true supercell reflections, which prevented us from finding the supercell propagation vector with any certainty. However, the off-centred positions of the supercell reflections from the main phase peaks in the single crystal data suggests a two-dimensional propagation vector, which would imply the non-equivalence of the *a* and *b* directions of the main phase, and might possibly remove the inversion centre. Two propagation vectors are required to fit the data: **k$_1$** = (-0.07, 0.13, 0) and **k$_2$** = (0.13, -0.07, 0). Since all of the observed supercell reflections so far occur at the same |**Q**|, it cannot be known whether or not these vectors are correct at the present time with absolute certainty. One interesting aspect is the apparent structure factor extinction of the intensities at (110) - **k$_i$** and (2-10) + **k$_i$**, which could aid in future single crystal neutron diffraction experiments to solve the complete supercell. Although the supercell reflections were only found in the neutron diffraction data, this may not be surprising if O ion displacements are responsible for it, which x-rays are not as sensitive to in this material. However, while oxygen nonstoichiometry may *influence* the supercell as manifested in the propagation vector (we did not have the necessary resolution to test this), the supercell is likely caused by a natural instability of the oxygen lattice rather than a deviation from the nominal stoichiometry. A number of samples were prepared under various conditions, and in all of the samples tested we either observed the supercell directly using neutron diffraction or observed no change in the bulk properties that we have associated with the supercell.

The observed supercell reflections remained invariant under all measured temperatures and magnetic fields within instrumental resolution. It is likely that the supercell has no clear role in the magnetoelectric properties of this material below the spin-flop transition; thus trying to control the oxygen stoichiometry is a not a viable route to tuning the magnetoelectric behavior. However, the supercell may indirectly validate a microscopic mechanism for enhanced magnetoelectric coupling; a permanent supercell formed by a periodic, perhaps incommensurate modulation of the $O^{2-}$ sublattice already takes care of most of the "displacive work" necessary for electric polarization under the application of a magnetic field. The movement of $Ti^{4+}$ from $Mn^{2+}$ due to magnetic exchange striction need only be very slight to induce a larger-than-expected response.

Above the spin-flop transition the magnetoelectric tensor acquires more obvious non-zero off-diagonal terms on account of the reduction of symmetry of the magnetic structure. [16, 17] Visually, the anomalies in the dielectric data are very different in the field regime above the spin flop transition than below it, indicating that an alternative mechanism for the observed polarization is likely at play. A transition from a collinear to noncollinear spiral structure similar to that proposed in Ref. [17] would result in a polarization along the *a*-axis via the Dzyaloshinskii-Moriya interaction. $Cr_2O_3$ shows similar behaviour, where asymmetric off-diagonal terms have been ascribed to ferrotoroidic order. [45]

Despite the complexity of the Mössbauer spectroscopy data, a number of observations can be made. Although previous studies have been made on Fe-doped $MnTiO_3$, the bulk of them were more focused on the intermediate members of the $FeTiO_3$-$MnTiO_3$ solid solutions that show spin behaviour that differs from both end series members. [5] By far the most comprehensive Mössbauer studies thus far on $MnTiO_3$ are that of Syono et al. [46] and the $^{119}$Sn Mössbauer studies performed by Fabritchnyi et al. [38] and Korolenko et al. [47] Syono et al. [46] doped 2 mol% $^{57}$Fe by reacting $Fe_2O_3$ with $TiO_2$ and pre-reacted $MnTiO_3$ under a reducing atmosphere to give a divalent Fe-containing material. This was verified in the isomer shift, found to be approximately 1.13 mm/s. A quadrupole splitting was found to be approximately Δ=0.85 mm/s at 300 K increasing to 1.5 mm/s by 4.2 K, however, the authors do not speak to the physical origin of a temperature dependent electric field gradient that is indicated by their fitted Δ. The fitted Δs are likely due to an elongation of the ligands along the c-axis. On the other hand, Fabritchnyi et al. [38] found no temperature dependence of the quadrupole splitting in the paramagnetic region on $Sn^{4+}$ doped $MnTiO_3$ samples, with much smaller values around 0.4 to 0.45 mm/s. The temperature dependence of the isomer shift was also consistent with a second-order Doppler shift effects. Below the $T_N$, three sites were used to fit the data with $B_{HF}$s of 52.5±0.5 T, 25±5 T, and 0 T, in very good agreement with our results. While the first and second sites were attributed to $Sn^{4+}$ on the Mn and Ti sites respectively, the third site was attributed to unreacted $SnO_2$ or $Ti_{1-x}Sn_xO_2$ clusters. Significant line broadening was also observed at the onset of the transition. Korolenko et al. [47] found that $^{119}$Sn ions impregnated on the surface of $MnTiO_3$ had drastically different spectra dependent on the crystal annealing conditions.

Our measurements yield results that are consistent with these other studies. In particular, three things can be learned from the Mössbauer spectra. Firstly, $Fe^{3+}$ occupies both of the $Mn^{2+}$ and $Ti^{4+}$ sites in roughly equal proportion. Secondly, it can be concluded that spin-lattice coupling is not as weak as originally proposed by Mufti et al. [16]: significant line broadening exists near $T_N$ that directly indicates relaxation phenomena. Even at 275 K, the two sites are broadened far beyond the intrinsic linewidth, but this could also be due to the presence of multiple, very similar sites arising from a long-range periodic modulation of the structure. Furthermore an anomalous variation is found in the f-factor at $T_N$, which can be correlated with the presence of magnons that soften as $T_N$ is approached from below. [8] Finally, while little is understood about the extra sites required to fit the lowest temperature data, unlike the Sn-study, no evidence for impurities were found from electron microprobe, magnetic susceptibility, heat capacity, or diffraction data, suggesting that these extra sites are intrinsic to the sample.

**5.0 – Conclusion**

$MnTiO_3$ was prepared and the previously reported electric and magnetic properties were verified. A new, previously unreported incommensurate supercell was found in four different powder and single crystal samples on two separate instruments. To within instrumental resolution, the peaks are both temperature and magnetic field independent from 300 to 1.8 K and 0 to 8 T. This suggests that the supercell likely stems from oxygen displacements. At the bulk level, it was found that an electric polarization flop occurs beyond the spin flop transition. On the other hand, Mössbauer spectroscopy was used to examine the possible influence of the supercell at the local level. Although the role of the supercell in relation to the multiferroic properties remains poorly understood, one likely microscopic explanation is local spin-lattice coupling, which can be used to resolve the data from the bulk vs the local probes.

**6.0 – Acknowledgements**

This work was supported by the National Sciences and Engineering Research Council of Canada (NSERC), The American Chemical Society Petroleum Research Fund, and Canada Foundation for Innovation. H.J.S. graciously thanks financial support from the Vanier Canada Graduate Scholarship (NSERC) and support from the University of Manitoba. H.J.S. also thanks R. Desautels for assisting with instrumentation and M. Bieringer for useful discussions. P.M.S. and C.M. would like to acknowledge support from the NSERC Canada Graduate Scholarship and Undergraduate Student Research Award respectively. C.R.W. is supported by the Canada Research Chair (Tier II) program and the Canadian Institute for Advanced Research. H.D.Z. is supported by the National Science Foundation (NSF-DMR-1350002). A portion of this research at Oak Ridge National Laboratory's High Flux Isotope Reactor was sponsored by the Scientific User Facilities Division, Office of Basic Energy Sciences, U.S. Department of Energy. A portion of this work was performed at the National High Magnetic Field Laboratory, which is sponsored by National Science Foundation Cooperative Agreement no. DMR-1157490, the State of Florida, and the U.S. Department of Energy. All of the authors are indebted to Ravinder Sidhu at the Department of Geological Sciences, University of Manitoba, and the Manitoba Institute for Materials for the use of their electron microprobe instrument. Finally, all of the authors would like to thank S. Cadogan for generously providing us with enriched iron.

**Table 1:** Selected refined values found for the 300 K joint refinement of the stoichiometric sample. The lattice constants were found to be $a$ = 5.13790(9) and $c$ = 14.2821(3) Å. All error values have been corrected to account for correlated residuals. For the x-ray pattern, $R_p$ = 4.11, $R_{wp}$ = 5.47 and $\chi^2$ = 1.82. For the neutron pattern, $R_p$ = 3.52, $R_{wp}$ = 4.34 and $\chi^2$ = 8.34.

| Atom | $x$ | $y$ | $z$ | $B_{iso}$ (Å$^2$) | B. V. S. |
|---|---|---|---|---|---|
| Mn | 0 | 0 | 0.3598(5) | 0.7(1) | 2.06 |
| Ti | 0 | 0 | 0.1479(5) | 0.30(17) | 3.99 |
| O | 0.3478(5) | 0.0450(3) | 0.0896(2) | 0.59(3) | |

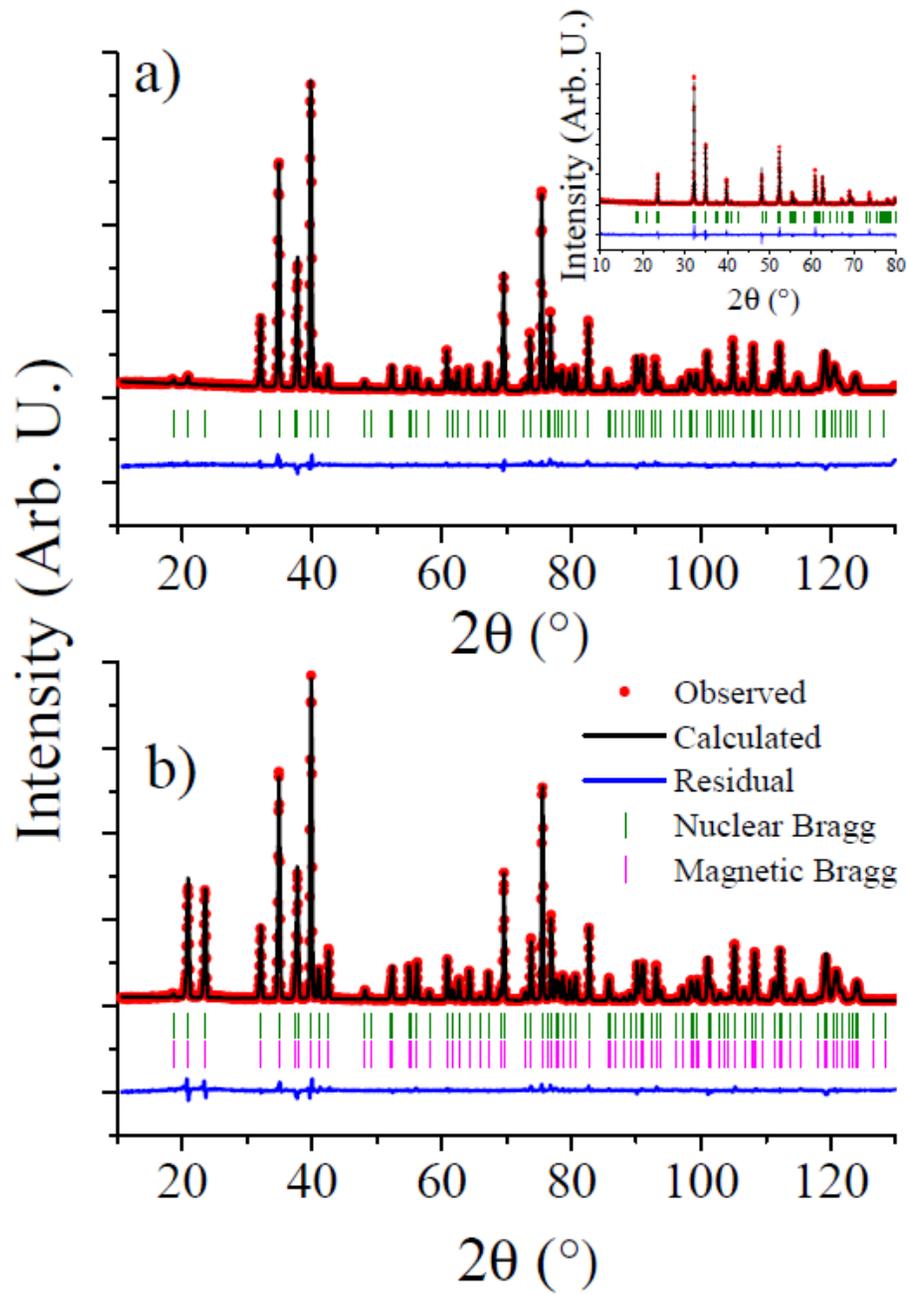

**Figure 1**: Powder neutron diffraction data collected on HB-2A for the stoichiometric sample at **a)** 300 K and **b)** 10 K. The inset depicts the corresponding x-ray diffractogram of the same sample used in the refinement.

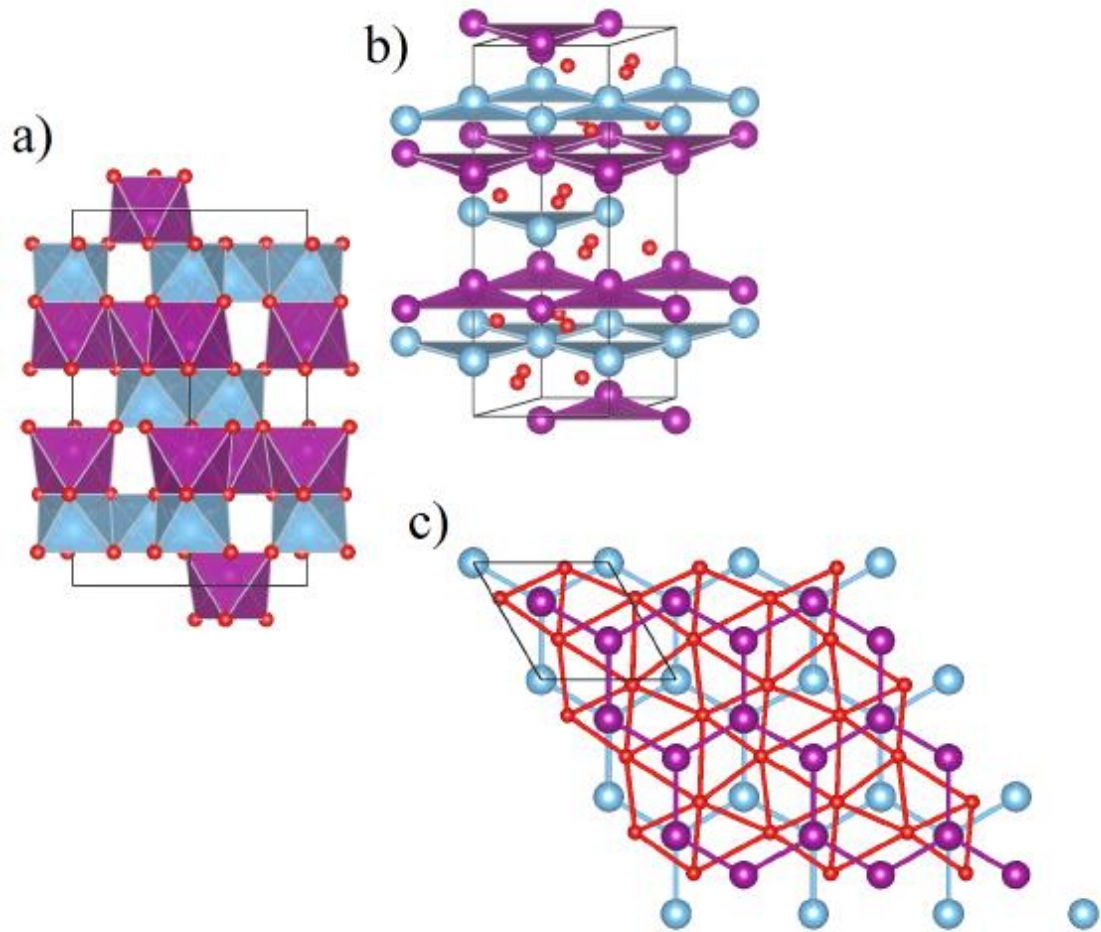

**Figure 2**: $Mn^{2+}$ (large dark, violet) and $Ti^{4+}$ (large light, blue) are octahedrally coordinated with $O^{2-}$ (small dark, red) and are arranged in buckled hexagonal layers, which are shown directly along [110] in **a)** and slightly offset from [100] in **b)**. The distorted close-packing of the anion network and trigonal cationic sublattices are emphasized in **c)** shown along [001].

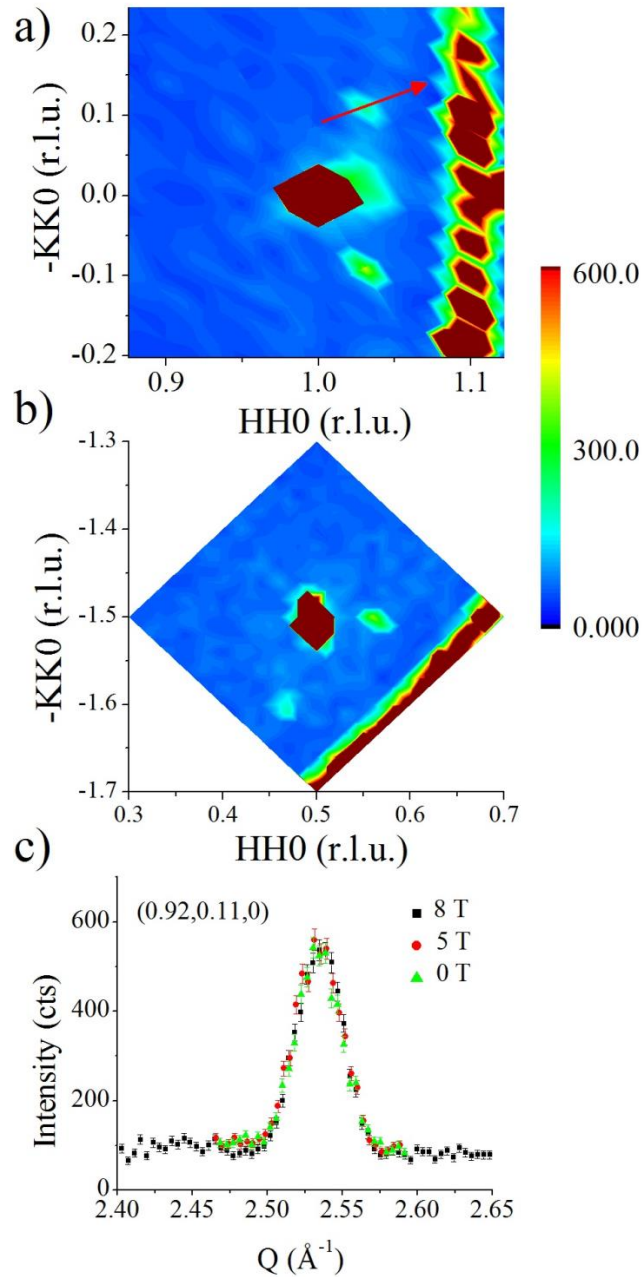

**Figure 3: a)** Mesh scan of the HK0 plane depicting the extra reflections flanking the (110) peak at 8 T. The 0 T mesh scan is unchanged (not shown). Extra regions of intensity are from the aluminum sample can; **b)** Mesh scan of the (2-10) reflection at 0 T. A small difference in intensity between the extra reflections is partly due to the broad step size in the mesh scan as well as the small misalignment (the crystal was aligned directly on the (110) reflection. The color scale used in both contour plots is in total counts over 15 seconds per point; **c)** Radial scan through the reflection denoted along the arrow shown in a) at $\mu_0 H$ = 8, 5, and 0 T. To give the reader an idea as to the relative intensities of these reflections, the (110) reflection yields 60 000 cts/s.

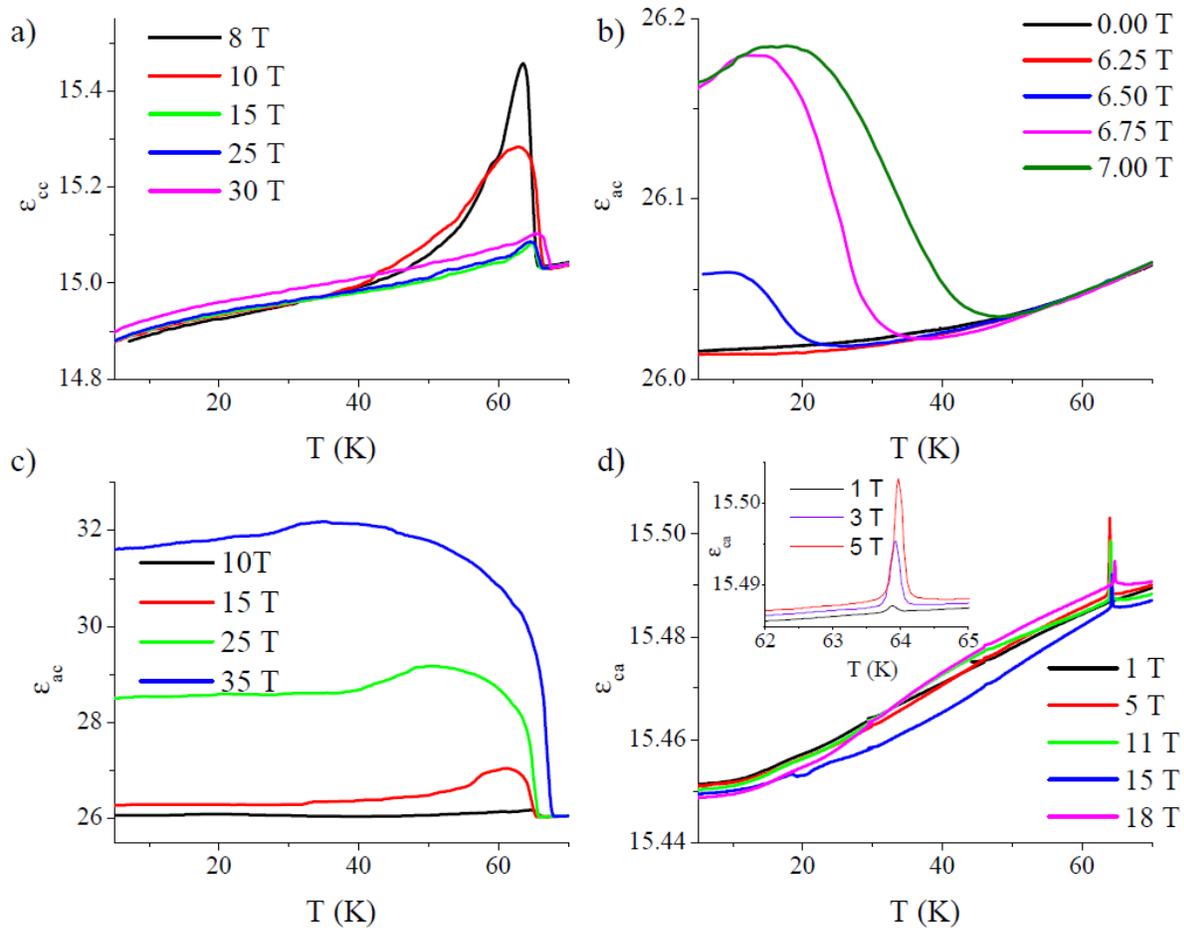

**Figure 4: a)** The dielectric constant with both magnetic and electric fields applied along *c*. As the magnetic field is increased further above µ₀H_SF, the anomaly merges with the broad feature that begins to develop just above 6.7 T. [23] By 30 T, only a small kink in the susceptibility remains. The dielectric constant as measured with the electric field now applied along the *a*-axis is shown in magnetic field applied along *c* **b)** below and **c)** above the spin flop transition. **d)** The dielectric constant was also measured with the magnetic field applied along *a* and the electric field along *c*.

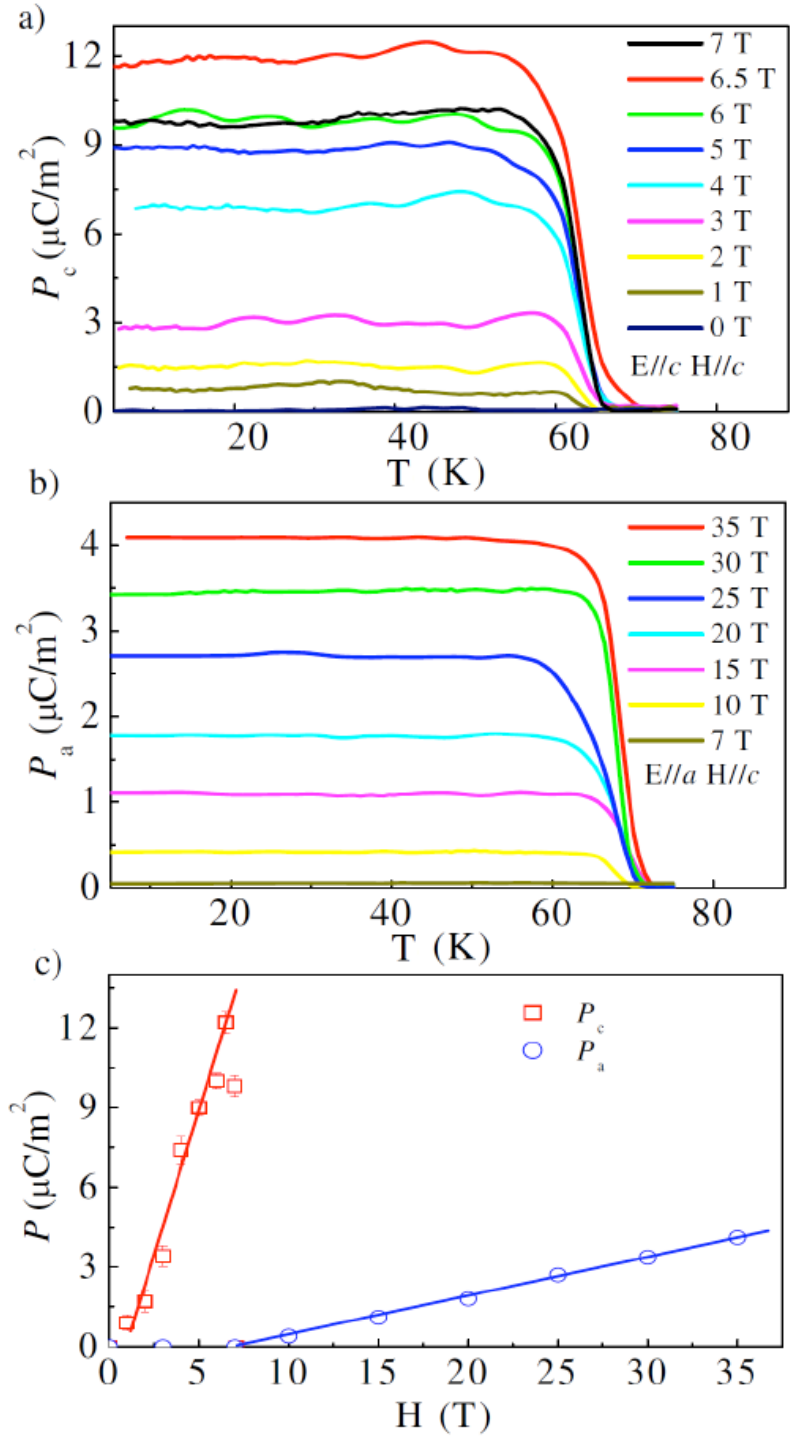

**Figure 5:** The polarization from the pyroelectric current measurements at various fields **a)** below and **b)** above the spin flop transition; **c)** A polarization flop is clearly observed accompanying the spin flop transition.

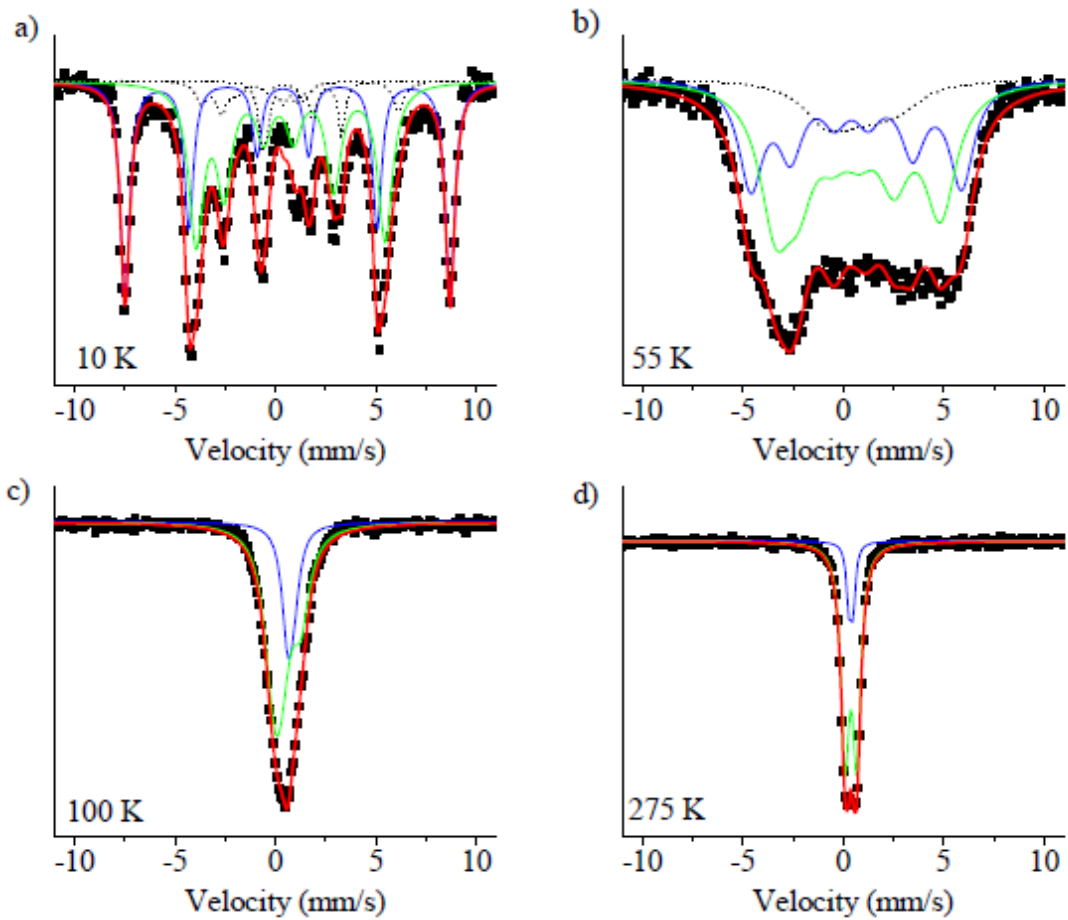

**Figure 6:** Fitted Mössbauer spectra at **a)** 10 K, **b)** 55 K, **c)** 100 K and **d)** 275 K. The blue and green curves represent the two major sites while the black dotted lines represent extra sites that cannot be well described at this time. The red curve is the overall fit to the data. Even at 275 K, well above $T_N$, the linewidth is still considerably broad.

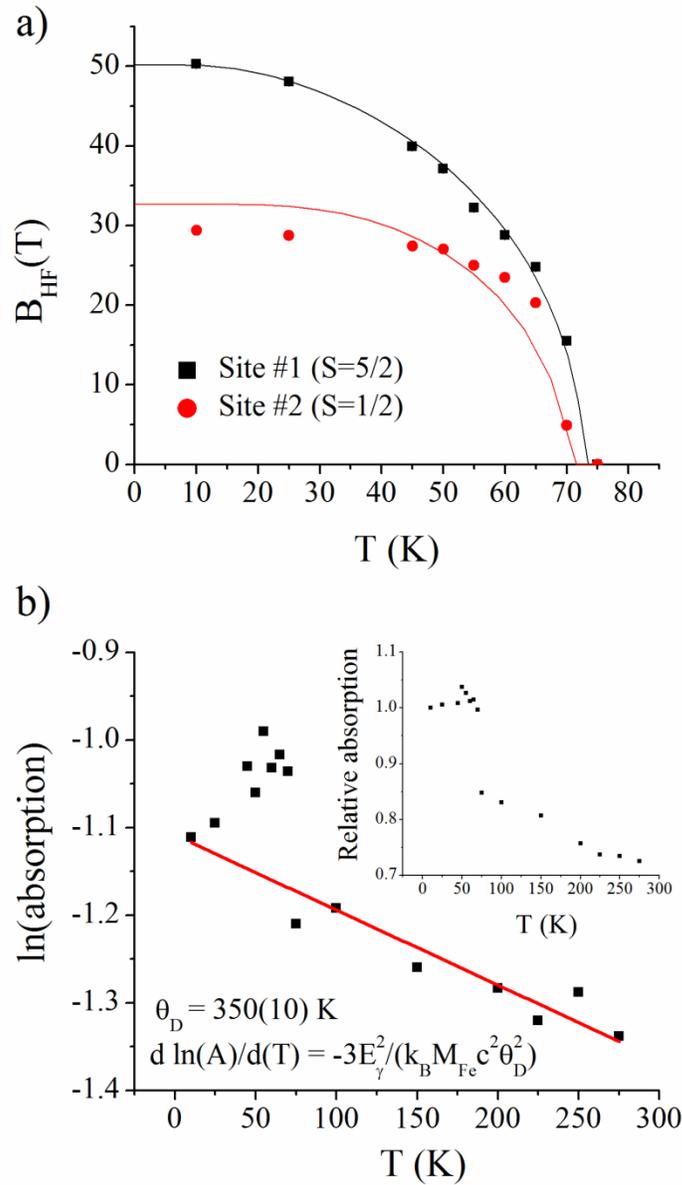

**Figure 7: a)** The temperature dependence of the hyperfine field for the two major sites (black squares and red circles) are fit using the Brillouin function (solid lines). One of the sites is well-modelled using the S=5/2 Brillouin function while the other is poorly modelled with the S=1/2 Brillouin function: likely an unphysical result; **b)** The *f*-factor is proportional to the spectral absorption and contains information about the phonon density of states over the first Brillouin zone. The red line is a linear fit, the slope of which is used to estimate the Debye temperature, $\theta_D$. $E_\gamma$ is the Mössbauer γ-ray energy, $k_B$ is the Boltzmann constant, $M_{Fe}$ is the atomic mass of $^{57}$Fe and *c* is the speed of light. The points at temperatures between 45 and 75 K were not included in the fit. This figure depicts the absorption found by fitting the spectra as temperature was increased step-wise on a log scale. The **inset** depicts the fits to the spectra as the temperature was decreased step-wise on a linear scale.